\documentstyle[12pt]{article}

\catcode`\@=11
\long\def\@makefntext#1{ 
\protect\noindent \hbox to 3.2pt {\hskip-.9pt
$^{{\ninerm\@thefnmark}}$\hfil}#1\hfill} 

\def\thefootnote{\fnsymbol{footnote}}
 \def\@makefnmark{\hbox to 0pt{$^{\@thefnmark}$\hss}}  

\def\ps@myheadings{\let\@mkboth\@gobbletwo
\def\@oddhead{\hbox{} 
\rightmark\hfil\ninerm\thepage}
\def\@oddfoot{}\def\@evenhead{\ninerm\thepage\hfil 
\leftmark\hbox{}}\def\@evenfoot{}
\def\sectionmark##1{}\def\subsectionmark##1{}}

\begin{document}

\newcommand{\symbolfootnote}{\renewcommand{\thefootnote}
	{\fnsymbol{footnote}}}
\renewcommand{\thefootnote}{\fnsymbol{footnote}}
\newcommand{\alphfootnote}
	{\setcounter{footnote}{0}
	 \renewcommand{\thefootnote}{\sevenrm\alph{footnote}}}

\newcounter{sectionc}\newcounter{subsectionc}\newcounter{subsubsectionc}
\renewcommand{\section}[1] {\vspace{0.6cm}\addtocounter{sectionc}{1}
\setcounter{subsectionc}{0}\setcounter{subsubsectionc}{0}\noindent
	{\bf\thesectionc. #1}\par\vspace{0.4cm}}
\renewcommand{\subsection}[1] {\vspace{0.6cm}\addtocounter{subsectionc}{1}
	\setcounter{subsubsectionc}{0}\noindent
	{\it\thesectionc.\thesubsectionc. #1}\par\vspace{0.4cm}}
\renewcommand{\subsubsection}[1]
{\vspace{0.6cm}\addtocounter{subsubsectionc}{1}
	\noindent {\rm\thesectionc.\thesubsectionc.\thesubsubsectionc.
	#1}\par\vspace{0.4cm}}
\newcommand{\nonumsection}[1] {\vspace{0.6cm}\noindent{\bf #1}
	\par\vspace{0.4cm}}

\newcounter{appendixc}
\newcounter{subappendixc}[appendixc]
\newcounter{subsubappendixc}[subappendixc]
\renewcommand{\thesubappendixc}{\Alph{appendixc}.\arabic{subappendixc}}
\renewcommand{\thesubsubappendixc}
	{\Alph{appendixc}.\arabic{subappendixc}.\arabic{subsubappendixc}}

\renewcommand{\appendix}[1] {\vspace{0.6cm}
        \refstepcounter{appendixc}
        \setcounter{figure}{0}
        \setcounter{table}{0}
        \setcounter{equation}{0}
        \renewcommand{\thefigure}{\Alph{appendixc}.\arabic{figure}}
        \renewcommand{\thetable}{\Alph{appendixc}.\arabic{table}}
        \renewcommand{\theappendixc}{\Alph{appendixc}}
        \renewcommand{\theequation}{\Alph{appendixc}.\arabic{equation}}
        \noindent{\bf Appendix \theappendixc #1}\par\vspace{0.4cm}}
\newcommand{\subappendix}[1] {\vspace{0.6cm}
        \refstepcounter{subappendixc}
        \noindent{\bf Appendix \thesubappendixc. #1}\par\vspace{0.4cm}}
\newcommand{\subsubappendix}[1] {\vspace{0.6cm}
        \refstepcounter{subsubappendixc}
        \noindent{\it Appendix \thesubsubappendixc. #1}
	\par\vspace{0.4cm}}

\def\abstracts#1{{
	\centering{\begin{minipage}{30pc}\tenrm\baselineskip=12pt\noindent
	\centerline{\tenrm ABSTRACT}\vspace{0.3cm}
	\parindent=0pt #1
	\end{minipage} }\par}}

\newcommand{\bibit}{\it}
\newcommand{\bibbf}{\bf}
\renewenvironment{thebibliography}[1]
	{\begin{list}{\arabic{enumi}.}
	{\usecounter{enumi}\setlength{\parsep}{0pt}
\setlength{\leftmargin 1.25cm}{\rightmargin 0pt}
	 \setlength{\itemsep}{0pt} \settowidth
	{\labelwidth}{#1.}\sloppy}}{\end{list}}

\topsep=0in\parsep=0in\itemsep=0in
\parindent=1.5pc

\newcounter{itemlistc}
\newcounter{romanlistc}
\newcounter{alphlistc}
\newcounter{arabiclistc}
\newenvironment{itemlist}
    	{\setcounter{itemlistc}{0}
	 \begin{list}{$\bullet$}
	{\usecounter{itemlistc}
	 \setlength{\parsep}{0pt}
	 \setlength{\itemsep}{0pt}}}{\end{list}}

\newenvironment{romanlist}
	{\setcounter{romanlistc}{0}
	 \begin{list}{$($\roman{romanlistc}$)$}
	{\usecounter{romanlistc}
	 \setlength{\parsep}{0pt}
	 \setlength{\itemsep}{0pt}}}{\end{list}}

\newenvironment{alphlist}
	{\setcounter{alphlistc}{0}
	 \begin{list}{$($\alph{alphlistc}$)$}
	{\usecounter{alphlistc}
	 \setlength{\parsep}{0pt}
	 \setlength{\itemsep}{0pt}}}{\end{list}}

\newenvironment{arabiclist}
	{\setcounter{arabiclistc}{0}
	 \begin{list}{\arabic{arabiclistc}}
	{\usecounter{arabiclistc}
	 \setlength{\parsep}{0pt}
	 \setlength{\itemsep}{0pt}}}{\end{list}}

\newcommand{\fcaption}[1]{
        \refstepcounter{figure}
        \setbox\@tempboxa = \hbox{\tenrm Fig.~\thefigure. #1}
        \ifdim \wd\@tempboxa > 6in
           {\begin{center}
        \parbox{6in}{\tenrm\baselineskip=12pt Fig.~\thefigure. #1 }
            \end{center}}
        \else
             {\begin{center}
             {\tenrm Fig.~\thefigure. #1}
              \end{center}}
        \fi}

\newcommand{\tcaption}[1]{
        \refstepcounter{table}
        \setbox\@tempboxa = \hbox{\tenrm Table~\thetable. #1}
        \ifdim \wd\@tempboxa > 6in
           {\begin{center}
        \parbox{6in}{\tenrm\baselineskip=12pt Table~\thetable. #1 }
            \end{center}}
        \else
             {\begin{center}
             {\tenrm Table~\thetable. #1}
              \end{center}}
        \fi}

\def\@citex[#1]#2{\if@filesw\immediate\write\@auxout
	{\string\citation{#2}}\fi
\def\@citea{}\@cite{\@for\@citeb:=#2\do
	{\@citea\def\@citea{,}\@ifundefined
	{b@\@citeb}{{\bf ?}\@warning
	{Citation `\@citeb' on page \thepage \space undefined}}
	{\csname b@\@citeb\endcsname}}}{#1}}

\newif\if@cghi
\def\cite{\@cghitrue\@ifnextchar [{\@tempswatrue
	\@citex}{\@tempswafalse\@citex[]}}
\def\citelow{\@cghifalse\@ifnextchar [{\@tempswatrue
	\@citex}{\@tempswafalse\@citex[]}}
\def\@cite#1#2{{$\null^{#1}$\if@tempswa\typeout
	{IJCGA warning: optional citation argument
	ignored: `#2'} \fi}}
\newcommand{\citeup}{\cite}

\def\fnm#1{$^{\mbox{\scriptsize #1}}$}
\def\fnt#1#2{\footnotetext{\kern-.3em
	{$^{\mbox{\sevenrm #1}}$}{#2}}}

\font\twelvebf=cmbx10 scaled\magstep 1
\font\twelverm=cmr10 scaled\magstep 1
\font\twelveit=cmti10 scaled\magstep 1
\font\elevenbfit=cmbxti10 scaled\magstephalf
\font\elevenbf=cmbx10 scaled\magstephalf
\font\elevenrm=cmr10 scaled\magstephalf
\font\elevenit=cmti10 scaled\magstephalf
\font\bfit=cmbxti10
\font\tenbf=cmbx10
\font\tenrm=cmr10
\font\tenit=cmti10
\font\ninebf=cmbx9
\font\ninerm=cmr9
\font\nineit=cmti9
\font\eightbf=cmbx8
\font\eightrm=cmr8
\font\eightit=cmti8

\newcommand{\bea}{\begin{eqnarray}}
\newcommand{\eea}{\end{eqnarray}}
\newcommand{\beq}{\begin{equation}}
\newcommand{\eeq}{\end{equation}}
\newcommand{\nnel}{\nonumber \\ {}}

\newcommand{\tselea}[1]{\label{#1}}
\newcommand{\tseleq}[1]{\label{#1}}
\newcommand{\tbib}[1]{\bibitem{#1}}
\newcommand{\tref}[1]{(\ref{#1})}
\newcommand{\tcite}[1]{\cite{#1}}


\newcommand{\auth}[1]{{ #1}}
\newcommand{\tseno}[1]{[#1]}
\newcommand{\tsetit}[1]{{\it #1}}
\newcommand{\journal}[1]{{\it{#1}}}
\newcommand{\vol}[1]{{\bf #1}}
\newcommand{\yr}[1]{(19#1)}
\newcommand{\NP}{\journal{Nucl. Phys.}}
\newcommand{\PR}{\journal{Phys. Rev.}}
\newcommand{\PRL}{\journal{Phys. Rev. Lett.}}
\newcommand{\PL}{\journal{Phys. Lett.}}
\newcommand{\AP}{\journal{Ann. Phys.}}
\newcommand{\JMP}{\journal{J. Math. Phys.}}

\newcommand{\tnote}[1]{}


\newcommand{\munr}{\mu_{\rm nr}}
\newcommand{\rmnr}{{\rm nr}}
\newcommand{\half}{\frac{1}{2}}
\newcommand{\gsim}{\raise.3ex\hbox{$>$\kern-.75em\lower1ex\hbox{$\sim$}}}


\begin{flushright}
Imperial/TP/95-96/3 \\
hep-ph/9510298 \\
15th October, 1995 \\
\end{flushright}


\centerline{\tenbf THE CONDENSED MATTER LIMIT OF}
\centerline{\tenbf RELATIVISTIC QFT
\footnote{Talk
given at the 4th Workshop on Thermal Field Theories and their Applications,
6th-10th August 1995, Dalian, China.
To appear in the proceedings, to be published by World Scientific.
This talk available via anonymous FTP from {\tt euclid.tp.ph.ic.ac.uk}
as  {\tt papers/95-6\_3.tex} and \LaTeX
source and postscript  versions available via WWW
from {\tt http://euclid.tp.ph.ic.ac.uk/Papers/index.html }.  }
}
\vspace{0.8cm}
\centerline{\tenrm T.S. EVANS\footnote{E-mail: {\tt T.Evans@IC.AC.UK};
WWW: {\tt http://euclid.tp.ph.ic.ac.uk/$\sim$time} }}
\baselineskip=13pt
\centerline{\tenit Theoretical Physics, Blackett Lab., Imperial College,}
\baselineskip=12pt
\centerline{\tenit  Prince Consort Road, London, SW7 2BZ, U.K.}
\vspace{0.9cm}
\abstracts{I study how to apply relativistic quantum field theory to
condensed matter systems.   The motivation for this is examined and
then two separate elements are considered.  First we identify the
precise relationship between relativistic and non-relativistic
fields.  Second we consider the need for a chemical potential and
how one includes this in static and dynamical calculations. }

\vspace{0.8cm}

\twelverm   
\baselineskip=14pt


\tnote{``tnotes'' enabled.  Comments like these will not appear in the
final version.}

\section{Introduction}

There has been considerable recent interest in condensed matter systems
amongst cosmologists\tcite{thexp,ST,RayParis}.
This is because it seems possible to apply
methods used in the early universe
to some condensed matter systems.  The latter
calculations
can be compared to laboratory experiments which in turn can give us further
confidence in the early universe results.  The table below
illustrates some of the links which I will discuss in the paper.
\typeout{DAL - table}
\begin{table}[htbp] 
\centering 
\begin{tabular}{r||c|c}
& Cosmology & Condensed Matter \\ \hline \hline
Theory used & Relativistic QFT & Non-Relativistic QFT \\ \hline
Parameter Ranges &
$| \mu | \ll T \sim m$ & $T \sim m - \mu = -\mu_{\rm nr}
\ll \mu \sim m$ \\ \hline
Example Theory & GUTS & Liquid ${}^3$He,${}^4$He \\ \hline
Example defects & U(1) Cosmic Strings & Vortices \\ \hline
Equation of Motion & $(\partial_t^2 + k^2 + m^2) \Phi =0 $ &
$(i\partial_t - \frac{k^2}{2m^2} - \munr ) \Psi = 0 $
\\ \hline
Fundamental Field &
$\Phi = \frac{1}{\surd 2 \omega} [ \Psi + \bar{\Psi}^\ast ] $ &
$\Psi = \sqrt{\frac{\omega}{2}}\Phi + \frac{i}{\sqrt{2
\omega}}\Pi^{*}$ \\
\mbox{ } &
$\Pi = i \surd (\frac{\omega}{2}) [ \Psi^* - \bar{\Psi} ] $ &
$\bar{\Psi} = \sqrt{\frac{\omega}{2}}\Phi^{*} + \frac{i}{\sqrt{2
\omega}}\Pi $
\end{tabular}
\tcaption{Analogies between Relativistic and Condensed Matter systems.}
\label{tnf}
\end{table}
In particular we note the theoretical work on cosmic strings, and
the links made\tcite{thexp} between this and the experiments\tcite{exp} on
vortices on liquid helium, both ${}^3$He and ${}^4$He.  For
example there is the recent work using out-of-equilibrium quantum
field theory on the creation of cosmic
strings\tcite{GR}.  In this  a single complex scalar field was used
and it was important to identify the correct physical dynamical
behaviors, especially the mass, to be used in the calculations.
It would be interesting to look at the condensed matter limit as
this can be considered as a toy model for liquid ${}^4$He.  However we
need to identify precisely what are the non-relativistic fields in
terms of the relativistic fields.

First we must specify what we meant by the relativistic and
condensed matter limits.  The energy scales involved in the
relativistic problems are such that a particle with typical thermal
kinetic energy is moving at relativistic speeds, so $T \gsim m$.
Then the Hamiltonian used is built around free particles with the
usual dispersion relation $\omega(k) = \surd (k^2 + m^2)$.  In more
practical terms, this usually means that thermal fluctuations are of
order the energy scale of the theory, so this is where there is
likely to be interesting physics e.g.\ phase transitions.  In many
cosmological problems, the density of relativistic particles, $n_+$,
and relativistic anti-particles, $n_-$, is large but
essentially the same.
In equilibrium we have
\beq
n_{\pm} (k) = \frac{1}{\exp \{ \beta ( \omega(k) \mp \mu ) \} } ,
\eeq
so this means that we have a tiny chemical potential,
\beq
\left| \frac{n_+ - n_-}{n_+ +  n_-} \right| \ll 1
\; \; \Rightarrow \;  \;
\frac{|\mu|}{m} \ll 1
\eeq
For instance at high temperatures in the early universe
the ratio of the baryon/anti-baryon asymmetry to the
baryon density is around $10^{-10}$.

The condensed matter limit is a non-relativistic limit so that $T
\ll m$ and for free particles the dispersion relation is $\omega(k)
\simeq m + k^2/(2m)$.  For instance in liquid Helium we are working at $T \sim
1$K
or less.  However, the relevant particles are the Helium atoms whose
masses are equivalent to $10^{13}$K.  The other aspect of the
condensed matter limit, and a further difference from typical
relativistic problems, is that there are essentially no anti-particles
so that
\bea
\frac{n_-(k)}{n_+(k)} \ll 1 &\Rightarrow& \mu \simeq m .
\eea
It is for this reason that condensed matter texts use a different definition
for chemical potentials.  Here I will denote this alternative by
$\munr$.  The relationship to the chemical
potential, $\mu$, normally used in relativistic texts, is given by
\bea
\munr &:=& \mu - m \gg - m .
\tselea{munrdef}
\eea

Note that for Fermi systems, $\mu_{\rm nr}$ has a direct physical meaning
in terms of the Fermi surface as
it is the Fermi energy at zero temperature and plays a similar role
at non-zero temperatures.  For Bosons with  no
condensation present, $|\mu|$ is strictly less than the physical
mass $m$ and it has no direct physical interpretation in terms of
some `Bose-surface'.  With
Bose-Einstein condensation, $|\mu|=m$ exactly and $\mu$ is no longer
even a free parameter.  The treatment of this case is more
intricate.  I will focus on Bosonic systems in this talk.

\newpage
\section{Extracting the non-relativistic fields}

In looking at relativistic and non-relativistic bosonic systems, we
note that the usual equations of motion for the free fields differ
by being second order and first order in time derivatives respectively;
\bea
(\partial_t^2 + k^2 + m^2) \Phi &=& 0  ,
\tselea{KG}
\\
(-i\partial_t + \frac{k^2}{2m^2} - \munr ) \Psi &=& 0  .
\tselea{nrKG}
\eea
Note \tref{nrKG} is the Gross-Pitaevskii equation rather than the
time-dependent Ginzburg-Landau equation\tcite{thexp}.
The latter has no factor
of $i$ in front of the time derivative but is generally thought by
condensed matter theorists to
be a better model for systems like ${}^4$He.
\tnote{$\partial_t \equiv -ip_0$ according to my N-point Green
function papers.}
This
suggests that in trying to express the condensed matter fields
$\Psi$ in terms of the relativistic fields we must split up the
relativistic Klein-Gordon equation into two first order (in time)
equations.  This is sometimes done when solving differential
equations numerically, when one would take $\Phi$ and $\partial_t
\Phi$ as independent variables.  So we will work with $\Phi$
and its conjugate variable $\Pi$ (`$\sim \partial_t \Phi$')
and start not from the usual Lagrangian path integral but from
the Hamiltonian version.  In this way, $\Pi$ remains available as
an independent variable which we can use, whereas it is integrated
out in reaching the Lagrangian formulation.
Thus a suitable generating functional to
consider is
\bea
Z[j, j^{*};\mu] &=&
\int_B {\cal D}\Pi^{*}{\cal D}\Pi{\cal D}\Phi^{*}{\cal D}\Phi
\exp \left( i \tilde{S}_{e}[\Pi, \Phi ;\mu]
+ i \int (j^{*}\Phi + j\Phi^{*}) \right) ,
\tselea{Zc}
\eea
where
\bea
B &\equiv& \{ \Phi(t)=\Phi(t-i \beta) \}
\\
\tilde{S}_{e}[\Pi, \Phi ;\mu] &=&
\int_{0}^{-i \beta}d t \left[ \int d^3x \;
\left( \Pi\partial_t{\Phi} +  \Pi^{*}\partial_t{\Phi}^{*}  \right)
- \left( H - \mu Q \right)  \right] ,
\tselea{Sc}
\\
H &=& \int d^{3}x [ \Pi^{*}\Pi +(\nabla\Phi^{*} )(\nabla\Phi )
 + m^2 \Phi^{*}\Phi + O(\lambda)]
\tselea{Hc}
\\
Q &=& i\int d^{3}x\; (\Phi^{*}\Pi^{*} - \Pi\Phi ).
\tselea{Qc}
\eea
and $O(\lambda)$ represents all cubic and higher order terms.
It is useful to think of $S_e$ as a matrix
\bea
\lefteqn{\tilde{S}_{e}[\Pi, \Phi ;\mu] } \nnel
&=&
\int_{0}^{-i \beta}d t \int d^{3}x
\left( \surd \omega \Phi^* , \frac{-i}{\surd \omega} \Pi  \right)
\left( \begin{array}{cc}
-\omega & i\partial_t + \mu  \\
i \partial_t + \mu & -\omega
\end{array} \right)
\left( \begin{array}{c} \surd \omega \Phi \\
\frac{i}{\surd \omega} \Pi^*
\end{array} \right)  .
\tselea{Sc2}
\eea

Clearly proceeding straight from here produces the usual relativistic
formalism.  So what we have to do is to replace the four fields,
$\Phi,\Pi$ and their conjugates, by four new ones.  We do this by
finding a  canonical transformation which leaves us with a temporal first
order action and which is diagonal in the new fields up to and
including the quadratic terms.  This is sufficient
to specify the new combinations to be
\bea
\Psi &=& \sqrt{\frac{\omega}{2}}\Phi + \frac{i}{\sqrt{2
\omega}}\Pi^{*}
\tselea{Psidef}
\\
\bar{\Psi} &=&
\sqrt{\frac{\omega}{2}}\Phi^{*} + \frac{i}{\sqrt{2 \omega}}\Pi
\tselea{Psibardef}
\eea
together with their complex conjugates\tnote{After this
talk was given, I was shown a footnote, citation 6,
of Haber and Weldon\tcite{HW} where these relations were also given.}.
This linear transformation also means
that the measure in the path integral does not pick up any strange
terms
\beq
{\cal D}\Pi^{*}{\cal D}\Pi{\cal D}\Phi^{*}{\cal D}\Phi =
{\cal D}\Psi^{*}{\cal D}\Psi{\cal D}\bar{\Psi}^{*}{\cal
D}\bar{\Psi} .
\eeq
Physically what has been achieved is easy to see in terms of free
fields as we have that
\beq
\Psi_{\rm free} \sim
\int d^3 \vec{k} \; a(\vec{k}) e^{-i \vec{k}.\vec{x} + i \omega t }
, \; \; \; \; \bar{\Psi}_{\rm free}  \sim  \int d^3 \vec{k} \; b(\vec{k})
e^{-i \vec{k}.\vec{x} + i \omega t }
\eeq
where $a(\vec{k}),b(\vec{k})$ annihilate particles and
anti-particles respectively.

Putting this all together we find that the generating functional now
looks like
\bea
Z[j,j^*;\mu] &=&  \int {\cal D}\Psi^{*}{\cal
D}\Psi{\cal D}\bar{\Psi^{*}}{\cal
D}\bar{\Psi}\exp\{ \tilde{S}_{\mu, \rmnr, e} \}
\tselea{Zpp}
\\
\tilde{S}_{\mu, \rmnr ,e} &=&  S_{0, {\rm nr}}[\Psi^{*},\Psi ;\mu, \omega]
+   S_{0, {\rm nr}}[\bar{\Psi}^{*},\bar{\Psi} ;-\mu, \omega] \nnel
&& \mbox{      } + S_{I,{\rm nr}}[\Psi^{*},\Psi,\bar{\Psi}^{*},\bar{\Psi}]
\\
\tilde{S}_{0, {\rm nr}}[\Psi^*,\Psi,\mu,\omega]
&=& \int_{0}^{-i\beta} d t \int
d^{3}x\,\Psi^{*}\biggl[i\frac{\partial}{\partial t} -
\omega (\nabla) + \mu \biggr]\Psi
\eea
on removing total derivatives and in the last expression doing a
small momentum expansion.
Thus the quadratic part of the
non-relativistic  fields $\Psi$ and $\bar{\Psi}$ has a global $O(2)
\otimes O(2)$ symmetry.  Particles and anti-particles are separately
conserved under this symmetry.
However typical interaction terms, contained in $S_{I,{\rm nr}}$, break this
symmetry, e.g.\
\beq
\lambda |\Phi^* \Phi |^2 = \frac{\lambda}{4 \omega^2}
(\Psi^* \Psi + \bar{\Psi}^* \bar{\Psi} + \bar{\Psi}^* \Psi +
\Psi^* \bar{\Psi} ) .
\eeq

Note that the  global $O(2) \otimes O(2)$
symmetry  of non-interacting $\Psi,\bar{\Psi}$ fields is a mixture of
the original Poincar\'{e} symmetries and the original global $O(2)$
of  the $\Phi$ fields.   Put another way, the $O(2) \otimes O(2)$
symmetry transformations written in terms of the $\Phi$ fields are
non-local transformations.  If we make the transformation
\beq
\Psi \rightarrow e^{i(\theta+\eta)} \Psi , \; \; \; \;
\bar{\Psi} \rightarrow e^{i(\theta-\eta)} \bar{\Psi}
\eeq
then this is equivalent to making the transformation
\beq
\left( \begin{array}{c} \Phi \\ \Pi^* \end{array} \right)
\rightarrow e^{i \theta}
\left( \begin{array}{cc}
\cos ( \theta ) & - \frac{1}{\omega(\nabla)} \sin ( \eta ) \\
\omega (\nabla ) \sin ( \eta ) & \cos ( \theta )
\end{array} \right)
\left( \begin{array}{c} \Phi \\ \Pi^* \end{array} \right)  .
\eeq

An alternative way of looking at this approximate  $O(2) \otimes
O(2)$ symmetry is to make a transformation to $\Psi ',\bar{\Psi}'$
defined in the same way as before except that we replace $\omega$ by
the mass $m$ in \tref{Psidef} and \tref{Psibardef}.  This version is
mentioned by Haber and Weldon\tcite{HW} (see the sixth
citation/footnote).  Then these
$\Psi ',\bar{\Psi}'$ fields show only an approximate $O(2) \otimes
O(2)$ symmetry, the symmetry being broken by terms of order
$\vec{k}/m$ as well as by the interactions in $S_I$ but the symmetry
is now a standard global symmetry in terms of the $\Phi$ and $\Pi$
fields.  Since the transformation from $\Phi$ to $\Psi '$ is global,
this approximate  $O(2) \otimes O(2)$ symmetry is also a approximate
global symmetry of the original $\Phi$ Lagrangian.  It is not
usually discussed because the terms which break the symmetry are
large for relativistic particles.

\tnote{Note how the effective mass gap is $m \pm \mu$ yet the kinetic term
$k^2/(2m)$ is still scaled by $m$.  Surely this means that this is
{\em not} going to give a useful condensed matter model.  As we will
see in the next section, the pole of the physical propagator is
still at $m$ and not at $m \pm \mu$ and this is why $m$ is still
controlling the kinetic term rather than $m \pm \mu$.\tnote{Is this
what we want in condensed matter systems? Check out some simple
condensed matter examples, e.g.\ in Landau-Lif. or Abrikosov et
al.\tcite{AGD} or Fetter-Walecka\tcite{FW}.}
}

So far everything has been exact, for any value of $\mu$ and $T$.
In practice for a condensed matter system $m \simeq \mu \gg T$.  The
low temperature means that we can make the non-relativistic
approximation
\bea
\tilde{S}_{0, {\rm nr}}[\Psi^*,\Psi ,\mu,\omega]
&\simeq&
\int_{0}^{-i\beta} d t \int
d^{3}x\,\Psi^{*}\biggl[i\frac{\partial}{\partial t} +
\frac{1}{2m}\nabla^{2} + \mu - m \biggr]\Psi
\eea
The high particle density, or equivalently the large chemical potential,
means that the mass gap term in the anti-particle $\bar{\Psi}$ free
Hamiltonian dominates as $m-\mu \sim T \ll m \sim \mu$.\tnote{The sources
connected to the anti-particles are also going to be set to zero as
Green functions involving anti-particles are of no physical
relevance in such situations.}   This means that the quadratic part
of the anti-particle Hamiltonian is to a good approximation $ 2m
\bar{\Psi}^* \bar{\Psi} $.  On integrating out the anti-particle
fields we find that we just set $\bar{\Psi}^*=\bar{\Psi}=0$ and so
the anti-particle fields just drop out.\tnote{Check this out.  What
if we didn't put $\mu$ into the Hamiltonian? Answer: The statistical
weights of the anti-particle poles kills their contributions, see
\tref{tgf}.}

This then leaves us with
\bea
\lefteqn{Z[j,j^*;\mu] \simeq Z_{\rmnr} =} \nnel
&& \int {\cal D}\Psi^{*}{\cal D}\Psi
\exp \{ \int_{0}^{-i\beta}d t \int
d^{3}x\,\Psi^{*}\biggl[i\frac{\partial}{\partial t }
+ \frac{1}{2m}\nabla^{2} - m + \mu \biggr]\Psi
+ S_{I,{\rm nr}}[\Psi^{*},\Psi] \}
\tselea{Znr}
\eea
where for $\lambda | \Phi^* \Phi|^2$ we have $S_I[\Psi^*,\Psi ]
=\lambda (\Psi^* \Psi)^2/(4m^2)$.
This means that we are now left with an approximate theory, \tref{Znr},
which also has an exact $O(2)$ symmetry.  In terms of the original
theory, this new $O(2)$ is broken by the terms of
$O(k/m),O((\mu-m)/m)$ which are dropped in reaching \tref{Znr}.
It is a different symmetry from the original exact global
$O(2)$ of the relativistic theory.  The latter mixed particles and
anti-particles, whereas \tref{Znr} does not.

There is one caveat.  The $\Phi$-fields are periodic in imaginary
time, but the $\Pi$-fields are not.  This means that the
$\Psi$-fields can only be approximately periodic, although when
\tref{Znr} is used exact periodicity is assumed.

The case of fermionic fields is rather different from this bosonic
case as the Dirac equation is already first order in time.  The
problem with fermions is rather that one must generate the
$\nabla^2/2m$ term from the first order $\vec{\gamma}.\vec{\nabla}$
term, and also go from four by four Dirac gamma algebra to the two
by two Pauli matrices.  Essentially in this case one must follow a
Foldy-Wouthuysen approach, but this will be considered elsewhere.

\section{Chemical Potential}

There are two approaches to the chemical potential in thermal field
theory, which are distinguished by the way they split the physical
information between initial conditions and microscopic dynamics -
see table \ref{methods}.
\typeout{DAL - table}
\begin{table}[htbp] 
\centering 
\begin{tabular}{c||c|c}
 & Microscopic Dynamics & Initial Conditions \\ \hline \hline
Method I & $H_e=H-\mu Q$ & $\Phi_e(t_0) = \Phi_e(t_0-i \beta)$ \\ \hline
Method II &  H & $\Phi(t_0) = e^{\beta \mu}\Phi(t_0-i \beta)$
\end{tabular}
\tcaption{Two Approaches to Chemical Potentials.}
\label{methods}
\end{table}

In Method I, the $H$ and $\mu Q$ terms of the density
matrix are merged into a single an effective Hamiltonian, $H_e$ which then
explicitly depends on the chemical potential $\mu$.
\beq
\rho = e^{-\beta (H-\mu Q) } = e^{ -\beta H_e }
\eeq
One of the great advantages of this method is that this new
effective theory looks just like a theory with no chemical
potential.  This makes the boundary conditions the usual simple
(anti-)periodic ones for bosonic (fermionic) fields so that all the
usual well established methods can be called upon.  Not surprisingly
this approach is very  popular\tcite{HW,Ka,BD}.  While the initial
picture is simpler, it is often the case that new complications due
to the presence of a chemical potential arise further down the road.
 All fields, and hence all Green functions, in Method I have a
subscript $e$ to denote that $H_e$ is being used.

In Method II, all aspects of the equilibrium density matrix are
treated in the same way.  Thus one encodes both the temperature and
the chemical potential in the boundary conditions, and the
Hamiltonian used is physical one with no information about the
density matrix included.  This leads to a rather more complicated
boundary condition, which for bosonic fields is of the form
\beq
\Phi(t_0) = e^{\beta \mu} \Phi(t_0-i \beta) .
\tseleq{muper}
\eeq
This makes matters more difficult than the zero chemical potential
case from the very start.  The advantage of this approach is that
the density matrix is often best thought of as  defining the initial
conditions.  The microscopic dynamics of the physical fields, as
described by $H$, has nothing to do with the initial conditions.
Thus keeping the Hamiltonian used free of contamination from the
initial conditions means that Method II is a very physical approach.
This is the approach used in the review of Landsman and van
Weert\tcite{LvW}.  It is also interesting to note that it
effectively the way that standard condensed matter
texts\tcite{AGD,FW} deal with chemical potentials associated with
fermionic degrees of freedom.

In fact for most out-of-equilibrium problems, the reverse is also
true, that is the initial conditions need have nothing to do with
the microscopic dynamics. Indeed in Closed Time Path approaches to
non-equilibrium problems\tcite{LvW,Go4} the initial conditions may
always be encoded in terms of some `temperature' and some
`Hamiltonian' as we can always write the density matrix as  $\rho =
\exp \{ - \beta_{init} H_{init} \}$ . However there is no reason why
this $H_{init}$ has anything to do the Hamiltonian which controls
the microscopic interactions and hence evolution of the system.

Of course in many practical situations the initial conditions are
related to some physical Hamiltonian.  Equilibrium is an extreme
case where the same $H$ appears in initial density matrix and
evolution.  However, Method I is still using a mathematical trick to
`simplify' matters but it is doing it at the expense of physical
clarity.  The discussion above suggests that problems with Method I
may occur when looking at dynamical problems.  This is because $H$
not $H_e$ really controls the microscopic evolution yet in Method I
$H_e$ is used for everything.  We will now
justify this assertion in detail by looking first at the propagators
which appear in both methods, i.e.\ we will solve simplest problem,
that of a single free complex scalar field.

We will start by following Method I\tcite{Ka,BD,BBD}, which
is to use real fields $\phi_{e1},\phi_{e2}$,  where
$\Phi_e=(\phi_{e1}+i \phi_{e2})/\surd 2$.\tnote{We will use
subscript $e$ on fields, propagators and actions calculated within
Method I.}   Thus the action, $S_{e}$, is of the form
\bea
{S}_{e} &=& \int dt \int d^3\vec{x}
\left[ \frac{1}{2} \left( \frac{\partial \phi_{e1}}{\partial
t} + \mu \phi_{e2} \right)^2 - \frac{1}{2} (\nabla \phi_{e1})^2 -
\frac{1}{2} m^2 \phi_{e1}^2 \right. \nnel
& & \mbox{       } \left. +
\frac{1}{2} \left( \frac{\partial \phi_{e2}}{\partial
t} - \mu \phi_{e1} \right)^2 - \frac{1}{2} (\nabla \phi_{e2})^2 -
\frac{1}{2} m^2 \phi_{e2}^2
+ O(\lambda) \right] \\
&=& \int dt dt'\;
\left(\phi_{e1}(t),\phi_{e2}(t)\right)_a
D_{ab}(t;t')
\left( \begin{array}{c}\phi_{e1}(t') \\
\phi_{e2}(t') \end{array} \right)_b
+ O(\lambda).
\tselea{Seffre}
\eea
The inverse matrix propagator, $D^{-1}$, is given
in energy-momentum coordinates, less an overall conservation delta
function, by
\bea
D^{-1}(k_0=2\pi i n/\beta,\vec{k}) &=&
\left(
\begin{array}{cc}
k_0^2 - \omega^2 +\mu^2 & 2 \mu i  k_0 \\
-2\mu i k_0              & k_0^2-\omega^2+\mu^2
\end{array}
\right)
\eea
where for notational simplicity we have used the imaginary time form
for the propagator.   This propagator is not diagonal in the one and
two fields  but it is easy to find the two eigenvalues of $D^{-1}$
which are $(k_0+\mu)^2-\omega^2$ and $(k_0-\mu)^2-\omega^2$.  This
suggests that there are four poles in the propagator\tcite{BD} at $k_0=
\pm\omega \pm \mu$.  However, we started with a single complex
scalar field so we expect only two poles.  What has happened?

The solution is to look more carefully at the diagonalised
Lagrangian.  What we find is that the eigenstates are $(1,i)$ and
$(1,-i)$ and hence the diagonalisation process has merely moved us
from real  fields $\phi_{e1},\phi_{e2}$ back to the original complex fields
$\Phi_e,\Phi_e^*$. This is not surprising as we know that the latter are
eigenstates of the charge operator, the real fields are not.  It
makes more sense in this case (where the symmetry is unbroken) to
stick with the $Q$ eigenstates.  If we write out the free action after
we have diagonalised \tref{Seffre}, we obtain
\bea
{S}_{e} &=& \frac{1}{2} \frac{1}{(2 \pi)^4}
\sum_n \int d^3\vec{k} \; (\Phi_e^*,\Phi_e)_a
\left(
\begin{array}{cc}
(k_0 + \mu)^2 - \omega^2 & 0 \\
        0              & (k_0-\mu)^2-\omega^2
\end{array}
\right)^{-1}_{ab}
\left( \begin{array}{c}\Phi_e \\ \Phi_e^* \end{array} \right)_b
\tselea{Seffc}
\\
&=& - \frac{1}{\beta V} \int_0^{-i\beta} dt dt'
\; \left(
\half \Phi_e^* (t) G_e^{-1}(t,t') \Phi_e(t') +
\half \Phi_e(t) H_e^{-1}(t,t') \Phi_e^*(t') \right)
\\
&=&-\frac{1}{\beta V} \int_0^{-i\beta} dt dt'
\; \Phi_e^* (t) G_e^{-1}(t,t') \Phi_e(t')
\tselea{Seffc2}
\eea
where the two propagators $G$ and $H$ are defined and related
through
\bea
i G_e(t,t') &=& {\rm Tr} \{ e^{\beta(H-\mu Q}
T [\Phi_e(t) \Phi_e^*(t')] \}
\tselea{ppsprop}
\\
i H_e(t,t') &=& {\rm Tr} \{ e^{\beta(H-\mu Q}
T [\Phi_e^*(t) \Phi_e(t') ] \}
= i G_e(t',t)
\\
G_e^{-1}(k_0,\vec{k})
&=& [ (k_0+\mu)^2-\omega^2] = H_e^{-1} (-k_0,\vec{k})
\eea
were all the times are time ordering are Euclidean
valued, and the energies are taken at $k_0=2 \pi n / \beta$ where $n$
is an integer.
Thus we see from \tref{Seffc} that what appears to be four poles is
actually the same two poles repeated a second time but with the time or
energy argument reversed, as \tref{Seffc2} shows.  In fact we can
obtain \tref{Seffc2} much quicker by working with $\Phi,\Phi^*$ from the
start.  This should not be surprising as these are the fields which
are the eigenstates of the charge operator.  Anyway there are only
two distinct poles as one expects given that we have a single
complex scalar field.  For $G_e$, \tref{ppsprop}, the poles are located at
\beq
k_0 = \pm \omega - \mu .
\tseleq{poleI}
\eeq

Now let us turn to Method II and compare the propagator found there
with \tref{ppsprop}.
Essentially all we have to do is solve the Klein-Gordon equation
subject to the weighted periodic boundary condition \tref{muper}.
Of course one needs other boundary conditions but these are provided
by the equal time commutation relations. \tnote{(see for example the
appendix of Evans et al.\tcite{EGR}).}  It is simple to look up the
answer\tcite{LvW} and the time-ordered
two-point Green function, $G_t$, is given by
\bea
\lefteqn{iG_t(t,x;t',x') =
{\rm Tr} \{ e^{\beta(H-\mu Q} T [\Phi(t,x) \Phi^*(t',x')]
\} }
\nnel
&=& \frac{1}{(2 \pi)^4} \int d^4k \; \exp \{ -ik_0t +i\vec{k}.\vec{x} \}
\left[ \frac{i}{k_0^2-\omega^2 + i \epsilon} + \right.
\nonumber \\
&& \; \; \; \; \left.
\frac{1}{\exp \{ \beta(k_0 - \mu) \} -1 }
\frac{2 \pi \delta(k_0-\omega)}{2 \omega}
+ \frac{1}{\exp \{ \beta(- k_0 - \mu) \} -1 }
\frac{2 \pi \delta(k_0 + \omega)}{2 \omega}  \right]
\tselea{tgf}
\eea
Thus we see clearly that in Method II there are only two poles and
they are at the same place as they are for a free field at zero
density and arbitrary temperature, namely at
\beq
k_0 = \pm \omega .
\tseleq{poleII}
\eeq
So in Method II we find that the poles of the
free propagator do not depend on any of the many-body parameters,
and are only dependent on the form of the quadratic part of the
Hamiltonian. Thus the position of the poles depends only on
microscopic physics which is described by the true physical
Hamiltonian.  On the other hand, the weighting or residue of the
poles does depend on the initial conditions.  Here the real-time
propagator shows us this physics very clearly.  We have a zero temperature
term which is a pure
quantum fluctuation contribution so has no explicit $\mu$ or $\beta$.
 Then there is one particle and one anti-particle pole, each weighted by
the appropriate statistical distribution.  These represent
statistical fluctuations due to the real particles in the heat bath.
This is why the particle pole has a much larger residue than the
anti-particle pole when there
are many particles but few anti-particles (large positive $\mu$).
So Method II leaves us with a picture which fits our physical
intuition.

However, we now have an apparent contradiction between the results of
Method I \tref{poleI} and Method II \tref{poleII}.  As the
investigation of the relationships between various Green functions
and the Green functions calculated in different
formalisms\tcite{TSEnpt} showed, it is imperative that we are
extremely careful in identifying the Green functions we are
studying.

It is immediately obvious that the difference encountered here
between Methods I and II is not a result of looking at different
types Green functions.  In the analysis of Method I we looked at
Euclidean two-point Green functions and, implicitly, their analytic
continuations, the retarded and advanced propagators. It is well
known though that the real parts of these two-point functions are
identical to the real parts of the time-ordered Green function,
which was considered when looking at Method II\tcite{LvW,AGD,FW}.
So we must go back to basics to find the source of the
contradiction.

As we are looking at a propagator in Method I, let us write down the
generating functional
\bea
Z_e[j_e^*,j_e]&=& {\rm Tr} \{ e^{-\beta(H-\mu Q)}
T [ e^{-i\int (j_e^*  \Phi_e +\Phi_e^* j_e)} ] \}
\\
&=& \sum_{\Phi(\vec{x})} \;
{}_{H_e}\langle \Phi,t_0-i\beta |
T e^{-i\int (j_e^*  \Phi_e +\Phi_e^* j_e)}
| \Phi , t_0 \rangle
\tselea{Zedef}
\eea
where we have used the unphysical effective Hamiltonian $H_e=H-\mu Q$
of Method I to evolve the bra.  This means that in order to
calculate $Z_e$, say by using path integrals, we have to evolve all
fields and states using $H_e$ not the physical $H$.  Thus all the
fields which appear in $Z_e$, or its derivatives, satisfy in the
Heisenberg picture
\beq
\Phi_e(t) = e^{i H_e t} \Phi(0) e^{-i H_e t} .
\tseleq{phiefft}
\eeq
The fields appearing in Method I are therefore
{\em not} the same for an arbitrary time
 as the physical fields $\Phi$, which satisfy in
the Heisenberg picture
\beq
\Phi(t) = e^{i H t} \Phi(0) e^{-i H t} .
\eeq
This is the reason we have been carefully labeling the fields used in
Method I with the $e$ subscript.  Likewise, the Green functions must
be distinguished.

There is however a simple link between the two fields, and we see that
from \tref{phiefft}
\bea
\Phi_e(t) &=& e^{i (H - \mu Q) t} \Phi(0) e^{-i (H_e-\mu Q) t}
\tselea{phiefft2}  \\
&=& e^{-i \mu Q t} e^{i H  t} \Phi(0) e^{-i H t} e^{i \mu Q t}
\tselea{phiefft3}  \\
&=& e^{-i \mu Q t} \Phi(t) e^{i \mu Q t}
\tselea{phiefft4}  \\
\Phi_e(t) &=& e^{i q \mu t} \Phi(t) .
\tselea{phiefft5}
\eea
where we have assumed that our field has a charge of $q$.  The same
connection is also noted elsewhere\tcite{AGD} but there  is no
discussion of Method II as a practical calculational tool for bosons.

Now, to make
a link between the Green functions, we need to make a connection between
the generating functionals in Method I and II.  So we choose the sources
such that
\beq
 j_e(t) = e^{-i \mu t}j(t) .
\tseleq{jejrel}
\eeq
as this then gives
\beq
Z[j^*,j] = Z_e[j_e^*,j_e] .
\tseleq{ZeZ}
\eeq
The generating functional in Method I, $Z_e$, was given in
\tref{Zdef} and that of Method II is
\bea
Z[j^*,j]&=& {\rm Tr} e^{-\beta(H-\mu Q)}
T e^{-i\int (j^*  \Phi +\Phi^* j)}
\\
&=& \sum_{\Phi(\vec{x})} \;
{}_{H}\langle \Phi,t_0-i\beta |
T e^{-i\int (j^*  \Phi +\Phi^* j)}
| \phi , t_0 \rangle
\tselea{Zdef}
\eea
This means that Method II generates {\em physical} Green functions
e.g.\
\bea
\lefteqn{\Gamma^{(2N)}(t_1,\ldots ,t_N;s_1,\ldots, s_N) =
\frac{\partial^{(2N)} Z}{\partial j^*(t_1) \ldots
\partial j^*(t_n) \partial j(s_1) \ldots \partial j(s_N)} } \\
&=& {\rm Tr} \left[ e^{-\beta (H-\mu Q)} T [\Phi(t_1) \ldots \Phi(t_N)
\Phi^*(s_1) \ldots \Phi(s_N) ] \right]
\tselea{npt}
\eea
where the $T$ indicates path ordering appropriate for whatever finite
temperature formalism is being considered.
Likewise we see that with Method I we generate thermal expectation
values but of the {\em unphysical} fields $\Phi_e,\Phi_e^*$, namely
\bea
\lefteqn{ \Gamma_e^{(2N)}(t_1,\ldots ,t_N;s_1,\ldots, s_N) =
\frac{\partial^{(2N)} Z}{\partial j_e^*(t_1) \ldots
\partial j^*(t_n) \partial j(s_1) \ldots  \partial j_e(s_N)} }
\\
&=& {\rm Tr} \left[ e^{-\beta (H-\mu Q)} T [\Phi_e(t_1) \ldots \Phi_e(t_N)
\Phi_e^*(s_1) \ldots \Phi_e(s_N) ] \right]
\tselea{npte}
\eea
Now we can use \tref{phiefft5}, \tref{jejrel} and \tref{ZeZ} to show
how the unphysical Green functions of Method I, $\Gamma_e$ of \tref{npte},
are linked to the physical ones of Method II, $\Gamma$ of \tref{npt}.
We find
\bea
\Gamma^{(2N)}(t_1,\ldots ,t_N;s_1,\ldots, s_N) &=&
e^{ - i \mu \sum_j(t_j-s_j ) }
\Gamma_e^{(2N)}(t_1,\ldots ,t_N;s_1,\ldots, s_N)
\tselea{nptrel}
\eea
In particular, we find that for the propagators we have that
\bea
D(t,s) &=&
{\rm Tr} \left[ e^{-\beta (H-\mu Q)}
T [\Phi(t) \Phi^*(s) ] \right]
\\
&=& e^{-i \mu (t-s) } D_e(t,s) \\
D(k_0) &=& D_e(k_0-\mu)
\tselea{DeDrel}
\eea
These relations apply to any approximation to the full Green functions
and in particular for the free field case.  This resolves the paradox
we found above, \tref{poleI} vs.\ \tref{poleII}.  Equation
\tref{DeDrel} tells us that the poles in Method I and Method II are
related by a simple shift of $\mu$ in the energy.

It is important to note from the relation between the two generating
functions that the partition function is identical in both cases
\beq
Z=Z[0,0]=Z_e[0,0]
\tseleq{ZZePart}
\eeq
Thus both methods give the same answer for the partition function
and so will give answers directly for all thermodynamic quantities.
Indeed, any static quantity, i.e.\ time independent
or equivalently zero energy quantity, can be extracted directly from either
method.  Only when one has time dependence or is looking at non-zero
energy does one need to use formulae such as \tref{nptrel} to move
between the two formalisms.

Finally we mention a complication with symmetry breaking, or
equivalently Bose-Einstein condensation.
This area has long been of interest\tcite{CCUK,HW,Ka,BD,FW} and has
had a lot of recent attention as a result of the experimental
observation of Bose-Einstein condensation in ``Quantum Atom''
systems\tcite{BECatom}.
So far we have been working in a
situation where the symmetry is unbroken, $\langle \Phi \rangle =0$.
However what happens if we are in a phase where this is not true?  In
Method I with periodic boundary conditions we proceed as usual and
shift the fields
\bea
\Phi_e (t, \vec{x}) &=& v_e + \Psi_e(t,\vec{x}) \\
\langle \Phi_e \rangle = v_e && \langle \Psi_e \rangle =0 ,
\eea
where $v_e$ is a constant in the usual way.
However in Method II such a simple approach is not compatible with the
more complicated boundary conditions \tref{muper}.  By using the link
between the fields in Methods I and II, \tref{phiefft5} we see that
we have to use\tnote{Is constant $v_e$ still the lowest energy
solution?  After all we have messed up the kinetic terms.  Try the
classical limit.}
\bea
\Phi (t, \vec{x}) &=& v(t) + \Psi(t,\vec{x}) \\
\langle \Phi_e \rangle = v(t) , && v(t) = e^{q \mu t} v_e ,\;\; \; \;
\langle \Psi_e \rangle =0 .
\eea

\section{Conclusions}

We have seen how to link the fields used in relativistic and
non-relativistic theories.  Equation \tref{phiefft5} gives this link.  Note
that we worked within Method I but the formula works for Method II as
well.

I have also discussed how to link the two approaches used when
including  chemical potentials.  In Method I, fields $\Phi_e$ are
used which evolve according to $H_e=H-\mu Q$.  This means that for
these fields, we are measuring the energy on a scale shifted by
$\mu$ relative to the standard relativistic field theory definition.
  Thus Method I is actually quite close to Method II.  Method I is
only `unphysical' in the sense that it is measuring its energies
from a zero which is shifted from the standard of relativistic field
theory. Method I is favoured by condensed matter physicists when
dealing with bosonic  systems\tcite{AGD}.\tnote{They often plot $n$
vs.\ $k$ and while $k$ is like energy, a true energy plot would have
the boundary between occupied and unoccupied energies at zero if they
used Method I!  So for fermions where they use Method II we are OK.}
It is also used in relativistic texts for calculations of partition
functions\tcite{Ka,BD,BBD}. It is easy to use for free energies
because of the equality \tref{ZZePart} between the partition
function of Method I, and that of method II.

However, for dynamical quantities there is a difference between
Methods I and II, summarised by \tref{nptrel}.  In this case, Method
II is clearer in my opinion as it sticks rigidly to the conventions
of relativistic field theory. In particular all energies are
measured relative to the same zero, which is the energy of the
vacuum state (in a simple unbroken case).   However the same
physical information can be obtained from Method II. The problem is
that energies are then measured relative to a zero shifted by $q \mu$ from that
of
Method I and the standard relativistic zero.  This means that fields
with different charges under the various symmetry groups involved
will have their zero's of energy shifted by different amounts!  Thus
in trying to understand the electroweak model one runs a serious
risk of getting confused about the energies involved of various
modes. So in discussions of dynamics, poles of propagators or
whatever,  in a relativistic field particle context but where Method
I is used, such as in Bernstein and Dodelson\tcite{BD}, one must be very
careful that one understands that the energy zero has been shifted.
Shifting your zero of energy does not mean that anything interesting
has happened physically!

However it is a matter of personal taste which method one chooses to
use.  The same physical information is contained in both approaches.
One must just be very careful when interpreting the information and be
very clear what energy zero one is working with.
Method I is closest to the way condensed matter physicists work with
bosonic systems, with
energies measured relative to the Fermi surface (which is equal to
$\mu$ at zero temperature). Method II means that we stick to the
conventions used amongst relativistic particle physicists.

\section{Acknowledgements}

I would like to thank the Royal Society for their support through a
University Research Fellowship.  I have benefited greatly from
discussions with G. Karra, A.J.Gill, R.J.Rivers and D.A.Steer, with
whom various aspects of the work discussed here, or extensions
thereof, have been developed at Imperial College.   I thank Dr.
F.C.Khanna for enlightening discussions and hospitality at the
Theoretical Physics Institute of the University of Alberta,
Edmonton, Canada. I would finally like to thank the organisers for
such an enjoyable meeting.

\typeout{--- references ---}

\end{document}